\documentstyle[a4,aps,epsf]{revtex}

\def\s{\sigma}

\def\up{\uparrow}
\def\dd{\downarrow}
\def\las{\langle}
\def\ras{\rangle}

\def\nn{\nonumber}
\def\si{\mbox{sin}}
\def\co{\mbox{cos}}

\def\kp{{\bf k_\parallel}}

\begin{document}
\title{Tunnel magnetoresistance in double spin filter junctions}
\author{Alireza Saffarzadeh$\thanks{E-mail: a-saffar@tehran.pnu.ac.ir}$\\
\it{Department of Physics, Tehran Payame Noor University, Fallahpour St.,
\\Nejatollahi St., Tehran, Iran}}
\date{\today}
\maketitle

\begin{abstract}
We consider a new type of magnetic tunnel junction, which consists of two
ferromagnetic tunnel barriers acting as spin filters (SFs), separated by
a nonmagnetic metal (NM) layer. Using the transfer matrix method and the
free-electron approximation, the dependence of the tunnel magnetoresistance
(TMR) on the thickness of the central NM layer, bias voltage and temperature
in the double SF junction are studied theoretically. It is shown that the TMR
and electron-spin polarization in this structure can reach very large values
under suitable conditions. The highest value of the TMR can reach 99\%.
By an appropriate choice of the thickness of the central NM layer, the degree of
spin polarization in this structure will be higher than that of the single SF
junctions. These results may be useful in designing future spin-polarized
tunnelling devices.
\end{abstract}

\section{\bf Introduction}

In the past few years, tunnel magnetoresistance (TMR) in magnetic tunnel
junctions consisting of two ferromagnetic metal (FM) electrodes separated by a
thin insulator (FM/I/FM) has attracted much attention due to its promising
applications in magnetic field sensors and magnetic random access memory
\cite{Moodera0,Ino}. The TMR in FM/I/FM junctions depends on the degrees of
electron spin polarization of the two FM electrodes \cite{Jul}, but the lack
of nearly perfectly spin-polarized current, and temperature dependence
of the polarization, have limited the TMR in these junctions
\cite{Meservey,Moodera1}. If we use half-metallic electrodes in the magnetic
tunnel junctions, which are completely spin-polarized at $T$=0 K due to
complications such as disorder and surface effects, half-metallicity is
destroyed and thus the spin polarization and the TMR are decreased
\cite{Pickett}. However, by exploiting the spin filtering phenomenon, one
can create 100\% spin polarization and obtain a giant TMR.

The spin filtering effect in a ferromagnetic semiconductor
(FMS) has been demonstrated dramatically in spin-polarized tunnelling
experiments \cite{Moodera2}.
In these experiments, using an Al/EuS/Al tunnel junction, Moodera {\it et al}
obtained 85\% spin polarization for tunnel electrons. In another study with EuSe
barrier junctions \cite{Moodera3} and in the presence of an external magnetic
field, they reported near 100\% spin polarization. More recently,
LeClair {\it et al} \cite{LeC1} from the combination of a spin filter (SF)
barrier and a FM electrode, obtained a large TMR in an Al/EuS/Gd tunnel
junction, which is a new method for spin injection into semiconductors
\cite{Fied}. When a FMS layer such as EuS is used as a tunnel barrier, due to
the spin splitting of the conduction band in the FMS, tunnelling electrons see
a spin-dependent barrier height. Thus the probability of tunnelling for one
spin channel will be much larger than the other, and a highly spin-polarized
current may result. Although the TMR using other SF barriers has also
been studied \cite{Sue,Nowak,LeC2}, the results have shown only little success.

The TMR in single and double SF junctions has also been investigated
theoretically in recent years. Based on the two band model and free-electron
approximation \cite{Slon}, Li {\it et al} \cite{Li} studied the tunnelling
conductance and magnetoresistance of FM/FMS/FM junctions. The results showed
that a decrease or increase in tunnelling current strongly depends on the
magnetization orientations both in the electrodes and in the FMS layer. In
another theory with double SF junctions, Worledge {\it et al} \cite{Worl}
studied the TMR of the NM (nonmagnetic metal)/FMS/FMS/NM junctions, using a
simple model, and
obtained a large magnetoresistance. In a recent paper, Wilczynski {\it et al}
\cite{Wil} investigated tunnelling in NM/FMS/NM/FMS/NM junctions in a sequential
tunnelling regime. They found a strong enhancement of the TMR with increasing
spin splitting of the barrier height in the ferromagnetic barriers.
However, they have not investigated the effect of electric field and
temperature on the tunnel currents and spin polarization. Thus, other aspects
of this structure, such as the voltage dependence of TMR, remain to be
explained.

In the previous paper \cite{Saffar2}, using the single-site approximation for
the NM/FMS/NM/FMS/NM double SF junction, we studied the effect of the thickness
of the central layer on the spin polarization of tunnel electrons at $T$=0 K.
We showed that the tunnelling spin polarization has an oscillatory behavior with
the thickness, which is due to the spin asymmetry of the reflection at the
FMS/NM interfaces.

In this paper, using the transfer matrix method and the free-electron
approximation, we have extended our previous work \cite{Saffar2} to
investigate the effect of the thickness of the central layer, temperature
and applied bias on the TMR for tunnelling through a double SF junction.
The paper is organized as follows. In section 2, the transfer matrix approach
of the spin-polarized tunnelling through the double SF junctions
is described. In section 3, the numerical results for the TMR and the spin
polarization of the tunnel currents are discussed. The
conclusions are summarized in section 4.

\section{\bf Model and formalism}
The system we consider here is a NM/FMS/NM/FMS/NM sandwich structure in the
presence of an applied bias $V_a$, which is depicted in Fig. 1. For simplicity,
we assume the FMS layers, which act as SFs, are made of the same
material, while the outer NM electrodes, which are semi-infinite, and the central
layer are made of the same metal. This structure is grown in the $x$ direction.
In this case, in a free-electron approximation for the spin-polarized
tunnelling electrons, the longitudinal part of the effective one-electron
Hamiltonian can be written as
\begin{equation}\label{H}
H_x=-\frac{\hbar^2}{2m_j^*}\frac{d^2}{dx^2}+U_j(x)+V_{j}^{\s} \ ,
\end{equation}
where $m^*_j$ ($j$=1-5) is the electron effective mass in the $j$th layer and
\begin{equation}\label{F}
U_j(x)=\left\{\begin{array}{cc}
0, & x<0 \ ,\\
-eV_ax/(L-c)+U_2, & 0<x<b\  ,\\
-eV_ab/(L-c), &  b<x<b+c\  ,\\
-eV_a(x-c)/(L-c)+U_4, & b+c<x<L\  ,\\
-eV_a, & x>L\  ,
\end{array}\right.
\end{equation}
where $U_2$ and $U_4$ are, respectively, the barrier heights of the left and
right FMS layer at above $T_{C}$, $b$ and $d$ are the barrier widths, $c$ is the
width of the middle NM layer and $L=b+c+d$. The third term in Eq. (\ref{H})
which is a spin-dependent potential, denotes the $s-f$ exchange coupling
between the spin of tunnelling electrons and the localized $f$ spins in the
$j$th FMS layer. Within the mean field approximation, $V_{j}^{\s}$ is
proportional to the thermal average of the $f$ spins, $\las S_z\ras$ (a 7/2
Brillouin function), and can be written as
$V_{j}^{\s}=-I\s\las S_z\ras$. Here, $\s=\pm 1$, which corresponds to
$\s=\up,\dd$, respectively and $I$ is the $s-f$ exchange constant in the
FMS layers.

The Schr\"odinger equation for a barrier layer under the influence of an
applied bias can be simplified by a coordinate transformation whose solution
is a linear combination of the Airy function Ai[$\rho(z)$] and its complement
Bi[$\rho(z)$] \cite{Abram}. Considering all five regions of the
NM/FMS/NM/FMS/NM junction shown in Fig. 1, the eigenfunctions of the
Hamiltonian (\ref{H}) with eigenvalue $E_x$ have the following forms:
\begin{equation}\label{psi}
\psi_{j,\s}(x)=\left\{\begin{array}{cc}
A_{1\s}e^{ik_{1}x}+B_{1\s}e^{-ik_{1}x}, & x<0 \ ,\\
A_{2\s}\mbox{Ai}[\rho_{2\s}(x)]+B_{2\s}\mbox{Bi}[\rho_{2\s}(x)], & 0<x<b\ ,\\
A_{3\s}e^{ik_{3}x}+B_{3\s}e^{-ik_{3}x}, & b<x<b+c\  ,\\
A_{4\s}\mbox{Ai}[\rho_{4\s}(x)]+B_{4\s}\mbox{Bi}[\rho_{4\s}(x)], & b+c<x<L\ ,\\
A_{5\s}e^{ik_{5}x}+B_{5\s}e^{-ik_{5}x}, & x>L\  ,
\end{array}\right.
\end{equation}
where the coefficients $A_{j\s}$ and $B_{j\s}$ are constants to be determined
from the boundary conditions and
\begin{equation}
k_{1}=\sqrt{2m_1^*E_x}/\hbar\  ,
\end{equation}
\begin{equation}
k_{3}=\sqrt{2m_3^*(E_x+eV_ab/(L-c))}/\hbar\  ,
\end{equation}
\begin{equation}
k_{5}=\sqrt{2m_5^*(E_x+eV_a)}/\hbar\  ,
\end{equation}
\begin{eqnarray}\label{rho}
\rho_{j,\s}(x)=\frac{x}{\lambda_{j}}+\beta_{j,\s}\ ,\qquad j=2,4\ ,
\end{eqnarray}
with
\begin{equation}
\lambda_j=\left[-\frac{(L-c)\hbar^2}{2m^*_jeV_a}\right]^{1/3}\  ,
\end{equation}
\begin{eqnarray}
\beta_{j,\s}=\left\{\begin{array}{cc}
\frac{(L-c)[E_x-U_j-V_j^\s]}{eV_a\lambda_j}, & j=2\ ,\\
\frac{(L-c)[E_x-U_j-V_j^\s-eV_ac/(L-c)]}{eV_a\lambda_j}, & j=4\ .
\end{array}\right.
\end{eqnarray}

Although the transverse momentum $\kp$ does not appear in the above notations,
the effects of the summation over $\kp$ will be considered in our calculations.

\subsection{Transmission coefficients}
By using the boundary condition such that the wavefunctions and their first
derivatives are matched at each interface point $x_j$, i.e.
$\psi_{j,\s}(x_j)=\psi_{j+1,\s}(x_j)$ and $(m^*_j)^{-1}[d\psi_{j,\s}(x_j)/dx]=
(m^*_{j+1})^{-1}[d\psi_{j+1,\s}(x_j)/dx]$, we obtain a matrix formula that
connects the coefficients $A_{1\s}$ and $B_{1\s}$ with the coefficients
$A_{5\s}$ and $B_{5\s}$ as follows:
\begin{eqnarray}
\left[\begin{array}{c}A_{1\s}\\B_{1\s}
\end{array}\right]
=M_{total}\left[\begin{array}{c}A_{5\s}\\B_{5\s}\end{array}\right]\  ,
\end{eqnarray}
where
\begin{eqnarray}\label{M}
M_{total}&=&\frac{k_{5}}{k_{1}}
\left[\begin{array}{cc}
ik_{1}& \frac{1}{\lambda_2}\frac{m_1^*}{m_2^*}\\
ik_{1}&-\frac{1}{\lambda_2}\frac{m_1^*}{m_2^*}
\end{array}\right]
\left[\begin{array}{cc}
\mbox{Ai}[\rho_{2\s}(x=0)]&\mbox{Bi}[\rho_{2\s}(x=0)]\\
\mbox{Ai}'[\rho_{2\s}(x=0)]&\mbox{Bi}'[\rho_{2\s}(x=0)]
\end{array}\right]\nn\\&&
\times\left[\begin{array}{cc}
\mbox{Ai}[\rho_{2\s}(x=b)]&\mbox{Bi}[\rho_{2\s}(x=b)]\\
\frac{1}{\lambda_2}\frac{1}{m^*_2}\mbox{Ai}'[\rho_{2\s}(x=b)]&
\frac{1}{\lambda_2}\frac{1}{m^*_2}\mbox{Bi}'[\rho_{2\s}(x=b)]
\end{array}\right]^{-1}\nn\\&&
\times\left[\begin{array}{cc}
\co(k_3c)&-\frac{m^*_3}{k_3}\si(k_3c)\\
\frac{k_3}{m^*_3}\si(k_3c)&\co(k_3c)
\end{array}\right]\nn\\&&
\times\left[\begin{array}{cc}
\mbox{Ai}[\rho_{4\s}(x=b+c)]&\mbox{Bi}[\rho_{4\s}(x=b+c)]\\
\frac{1}{\lambda_4}\frac{1}{m^*_4}\mbox{Ai}'[\rho_{4\s}(x=b+c)]&
\frac{1}{\lambda_4}\frac{1}{m^*_4}\mbox{Bi}'[\rho_{4\s}(x=b+c)]
\end{array}\right]\nn\\&&
\times\left[\begin{array}{cc}
\mbox{Ai}[\rho_{4\s}(x=L)]&\mbox{Bi}[\rho_{4\s}(x=L)]\\
\mbox{Ai}'[\rho_{4\s}(x=L)]&\mbox{Bi}'[\rho_{4\s}(x=L)]
\end{array}\right]^{-1}\nn\\&&
\times\left[\begin{array}{cc}
ik_{5}&\frac{1}{\lambda_4}\frac{m_5^*}{m_4^*}\\
ik_{5}&-\frac{1}{\lambda_4}\frac{m_5^*}{m_4^*}
\end{array}\right]^{-1}
\left[\begin{array}{cc}
e^{-ik_{5}L}&0\\
0&e^{ik_{5}L}
\end{array}\right]^{-1}\  .
\end{eqnarray}

Since there is no reflection in region 5, the coefficient $B_{5\s}$ in
Eq. (\ref{psi}) is zero. In this case the transmission coefficient of the
spin $\s$ electron which is defined as the ratio of the transmitted flux to
the incident flux, for the double SF structure shown in Fig. 1, can be
written as
\begin{equation}\label{Ps}
T_\s(E_x,V_a)=\frac{k_{5}m_1^*}{k_{1}m_5^*}\left|\frac{1}
{M^{11}_{total}}\right|^2\ ,
\end{equation}
where $M^{11}_{total}$ is the left-upper element of the matrix $M_{total}$
which is defined in Eq. (\ref{M}).
Note that the transmission coefficient depends on the longitudinal energy
$E_x$, the applied bias $V_a$ and the spin orientation.
\subsection{Spin polarization and TMR}
The spin-dependent current density for single or double SF junctions at a given
applied bias $V_a$ can be calculated within the free-electron model \cite{Duke}:
\begin{equation}\label{j}
J_\s=\frac{em^*_1k_BT}{4\pi^2\hbar^3}\int_0^{\infty}T_\s(E_x,V_a)
\ln\left\{\frac{1+\exp[(E_F-E_x)/k_BT]}{1+\exp[(E_F-E_x-eV_a)/k_BT]}\right\}
dE_x\ ,
\end{equation}
where $k_B$ is the Boltzmann constant, $T$ is the temperature, and $E_F$ is the
Fermi energy.

The degree of spin polarization for the tunnel current is defined by
\begin{equation}\label{p}
P=\frac{J_\up-J_\dd}{J_\up+J_\dd}\ ,
\end{equation}
where $J_\up$ $(J_\dd)$ is the spin-up (spin-down) current density.
For the double SF junction, this quantity can be
obtained when the magnetizations of two FMS layers are in parallel
alignment.

The tunnel conductance per unit area is given by $G=\sum_{\s}J_{\s}/V_a$. In
this case, the TMR can be described quantitatively by the relative
conductance change as
\begin{equation}\label{tmr}
\mbox{TMR}=\frac{G_{\up\up}-G_{\up\dd}}{G_{\up\up}},
\end{equation}
where $G_{\up\up}$ and $G_{\up\dd}$ correspond to the conductances in the
parallel and antiparallel alignments of the magnetizations in the FMS layers,
respectively.

\section{\bf Numerical results and discussion}

The numerical calculations have been performed for a NM/EuS/NM/EuS/NM double
SF junction in which, for simplicity, we assume that the EuS layers have the
same thickness $b=d$=0.5 nm. The appropriate parameters for EuS which have been
used in this paper are: $T_C$=16.5 K \cite{Baum}, $S$=7/2, $I$=0.1 eV
\cite {Nolting}, $m^*_{\rm EuS}=1.5$ m$_e$ \cite{Wachter} and
$U_2=U_4=E_F$+0.75 eV. In the NM layers, the electron effective mass and
the Fermi energy are taken as $m^*_{\rm NM}=m_e$ and $E_F$=1.25 eV.
Here $m_e$ is referred to the free-electron mass.
In this study, we calculate the spin currents, tunnelling spin polarization
and the TMR by using Eqs. (\ref{j})-(\ref{tmr}), respectively. In our
considered system, the magnetization orientation (i.e. the $f$ spins'
direction) in the left EuS layer stays fixed but the other EuS layer is free
and may be switched back and forth by an external magnetic field (see Fig. 1).
Thus, when the magnetizations of two FMS layers are parallel, spin-up and
spin-down electrons see a symmetric structure, while for the antiparallel
alignment these electrons see an asymmetric structure. This structural
asymmetry results from the difference in the two barrier heights for each spin
channel.

Figure 2 shows the TMR as a function of the thickness of the central NM
layer at $T$=0, 0.45 and 0.9 $T_C$, when the bias voltage $V_a$=50 mV is
applied to the junction. It is obvious that the TMR oscillates with increasing
thickness $c$ and have well-defined peaks in which the TMR can reach 99\%
in some structures. The height of these peaks decreases with increasing $c$.
The oscillatory behavior is related to the quantum-well states of the central
NM layer and the spin-polarized resonant tunnelling. On the other hand, due to
the temperature dependence of spin splitting in the EuS layers, the barrier
heights become spin-dependent, so that with decreasing temperature, this
spin splitting, and thus the TMR, increases.

To understand the physical origin of the TMR and the oscillations, we study
the energy dependence of the transmission coefficients through the double SF
junction. In Fig. 3, we have plotted the spin-dependent transmission
coefficients at $T$=0 K for $c$=0.50 nm, which corresponds to a flat area
between the peaks, and for $c$=0.72 nm, which corresponds to a local maximum in
the TMR. Because of the quasibound states in the central NM layer, the
transmission coefficients reach unity at the resonance peaks which become
sharper in the low incident energy region, since in this energy region
the resonance levels are more strongly quantized.
The results for $c$=0.50 nm show that there is one resonance level in the
quantum well for both spin orientations and magnetic alignments. All these
resonance levels are far from the Fermi energy. However, the transmission
coefficient for spin-up electrons in the parallel alignment is higher than the
antiparallel alignment, and for spin-down electrons in both alignments.
Thus, there is a small difference in the current density $J$ (=$J_\up+J_\dd$)
in both magnetic configurations, which gives rise to relatively small TMR at
this thickness of the central layer. For $c$=0.72 nm the resonance states shift
to the lower energy side and a new resonance level slightly below $E_F$
appears, which is active only for spin-up electrons in the parallel alignment.
Therefore, there is a large difference in the current density $J$ in both
alignments and consequently gives rise to large TMR.
It is clear from the Fig. 3(b) and 3(d) that, for each thickness $c$, the
resonance levels for spin-up and spin-down electrons in the antiparallel
alignment coincide. The reason is that, for the antiparallel alignment,
electrons with up (down) spin are easy (difficult) to tunnel into the central
NM layer, and difficult (easy) to tunnel out of it; thus, both the spin-up and
spin-down electrons see the same effective height of the barriers during the
tunnelling process through the whole system.

In Fig. 4 we show the TMR as a function of the applied bias at $T$=0 K for the
thicknesses $c$ correspond to Fig. 3. With increasing bias voltage,
the TMR for $c$=0.50 nm decreases very slowly because, in this range of the
applied voltage, the discrepancy between the conductance for the parallel
alignment and that for the antiparallel alignment increases only slightly. On
the other hand, the results show that, for $c$=0.72 nm at the beginning, the
TMR slowly decreases with increasing the bias voltage. However, for the
voltages higher than $V_a$=80 mV, it quickly decreases. The reason is that, at
higher voltages, one of the resonance levels becomes active for both magnetic
alignments which drastically reduces the TMR.

It should be noted that, for very low values of the applied bias and the
incident energy, a numerical instability is occurred in some of our
calculations, which is due to the use of exact Airy functions. This
instability is overcome by using the asymptotic forms of Airy functions
\cite{Abram} and numerical analytical methods.

In Fig. 5 we have shown the spin polarization of the tunnelling current versus
normalized temperature $T/T_C$ for single and double SF junctions to reveal
the SF effects of the FMS layers from another point of view.
At high temperatures $T>T_C$, there is no spin splitting $E_{ex}$ in the
conduction band of the EuS layers and the transmission coefficients for two
spin channels coincide. Thus there is no TMR and spin polarization effect.
As the temperature decreases from $T_C$, the barrier heights for spin-up
electrons are lowered, while they are raised for spin-down electrons. This
temperature dependence of the barrier height, which is attributed to the
exchange splitting of the EuS conduction band, greatly increases the
tunnelling probability for one spin channel and reduces it for the other. In
the parallel alignment, the tunnel current for spin-up electrons is much
higher than the spin-down ones, which gives rise to the TMR and spin
polarization effect. On further decreasing the temperature, this
spin splitting and hence the difference in the barrier heights increases.
Therefore, the TMR and spin polarization reach the highest values at
$T$=0 K. The highest value of the spin polarization for the single SF junction
can reach 77\%, which is qualitatively in agreement with the experimental
measurements \cite{Moodera2} and the theoretical results
\cite{Metzke,Saffar1}, while for the double SF junction (in the parallel
alignment), it approaches 66\% for $c$=0.50 nm and 99\% for $c$=0.72 nm
\cite{Saffar2}, which is due to the change in the positions of the
spin-polarized quasibound states in the quantum well. Therefore, one can see
that, for the double SF junctions, the TMR and the spin polarization of
the tunnelling current can be controlled by the thickness of the central NM layer,
the temperature, and the applied bias.

In this study, the numerical results obtained for the TMR and the degree of
spin polarization can be compared with the result of Wilczynski {\it et al}
\cite{Wil}. As we discussed above, the enhancement of spin splitting $E_{ex}$
in the conduction band of the FMS layers can be obtained, when the temperature
decreases. In this case, the TMR and the spin polarization increases, and this
behaviour is in agreement with the result obtained by Wilczynski {\it et al} in
the sequential tunnelling regime.

\section{\bf Summary}

Based on the free-electron model, we presented a transfer matrix method for
spin-polarized tunnelling through the double SF junctions. The effect of the
quantum size, applied bias and temperature on the spin filtering and the spin
transport process in the FMS/NM double junctions are examined. Numerical
results indicate that the TMR oscillates as the thickness of
the central NM layer increases. It is further confirmed that, for some
thicknesses of the central NM layer, the spin-polarized resonant tunnelling
can gives rise to large values for the spin polarization and the TMR, even
at high temperatures ($T<T_C$). Therefore, in the system presented here,
it is able to select an appropriate applied bias, temperature and the
thickness to achieve a maximum TMR and spin polarization.

Although the temperature for observing a 99\% TMR is very low and the findings
of the paper are not directly applicable to spintronics technology, the results
may be useful in designing future spin-polarized tunnelling devices \cite{Dietl}.

In the present model, we have neglected the generally important complications,
such as the interface roughness, electron-electron interaction,
magnetic-domain wells, $f$ spin correlation, etc. These effects can play
important roles in the spin transport process and reduce the efficiency of spin
filtering.

\begin{figure}
\begin{center}
\leavevmode\hbox{\epsfxsize=1.\textwidth\epsffile{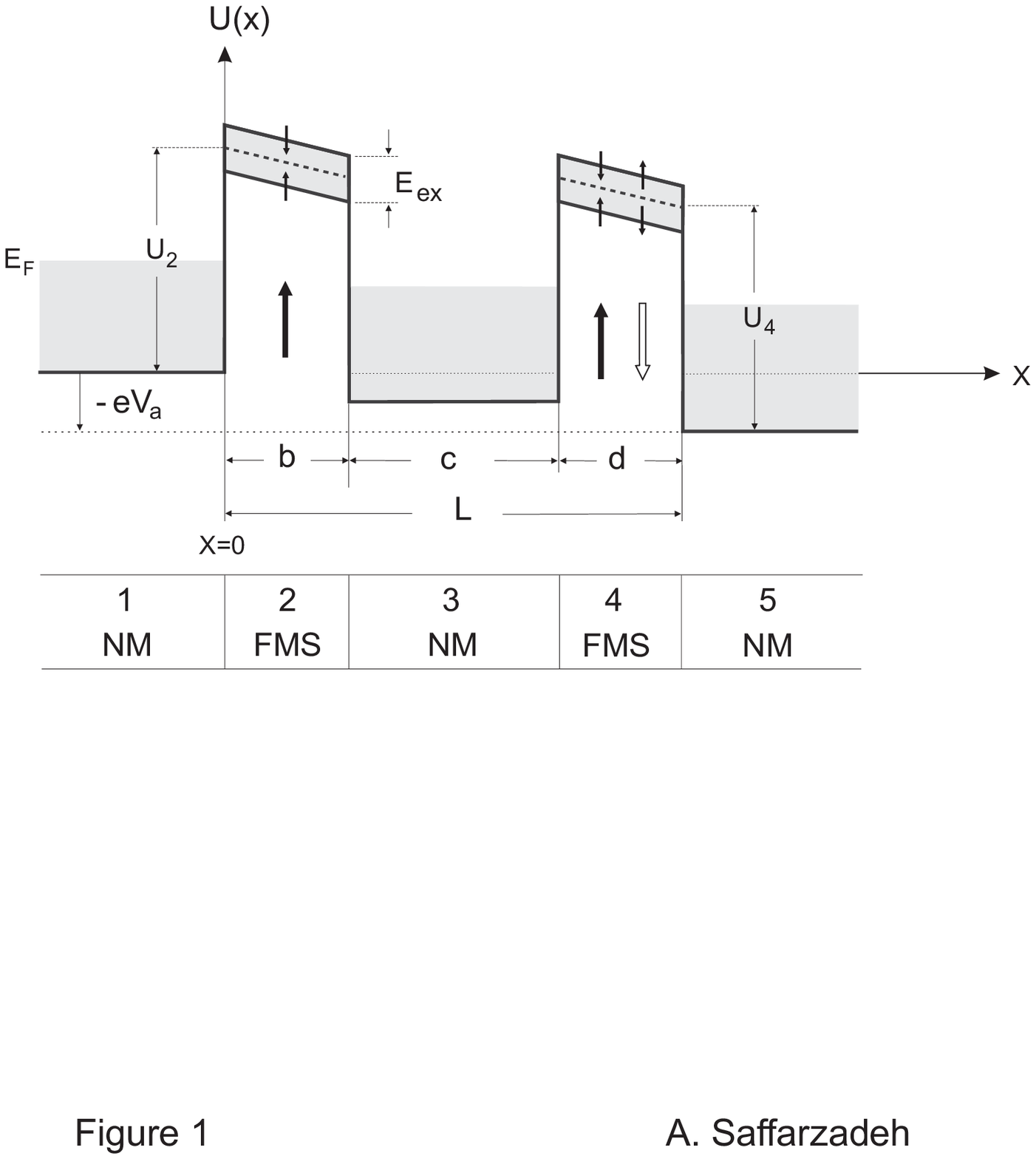}}
\end{center}
\caption{Spin-dependent potential profile for NM/FMS/NM/FMS/NM tunnel junctions
in the presence of a positive bias $V_a$.
The broken line in the FMS layers represents the bottom of the conduction
band at $T\geq T_C$. Below $T_C$, due to the exchange splitting $E_{ex}$ in the
conduction band, the barrier heights become spin-dependent, as indicated by
the thin arrows for spin-up and spin-down electrons.
The direction of magnetization in the left FMS layer is fixed in the $+z$
direction, while the magnetization in the right FMS
layer is free to be flipped into either the $+z$ or $-z$ direction, as
indicated by the filled and hollow arrows, respectively.}
\end{figure}
\newpage

\begin{figure}
\begin{center}
\leavevmode\hbox{\epsfxsize=1.1\textwidth\epsffile{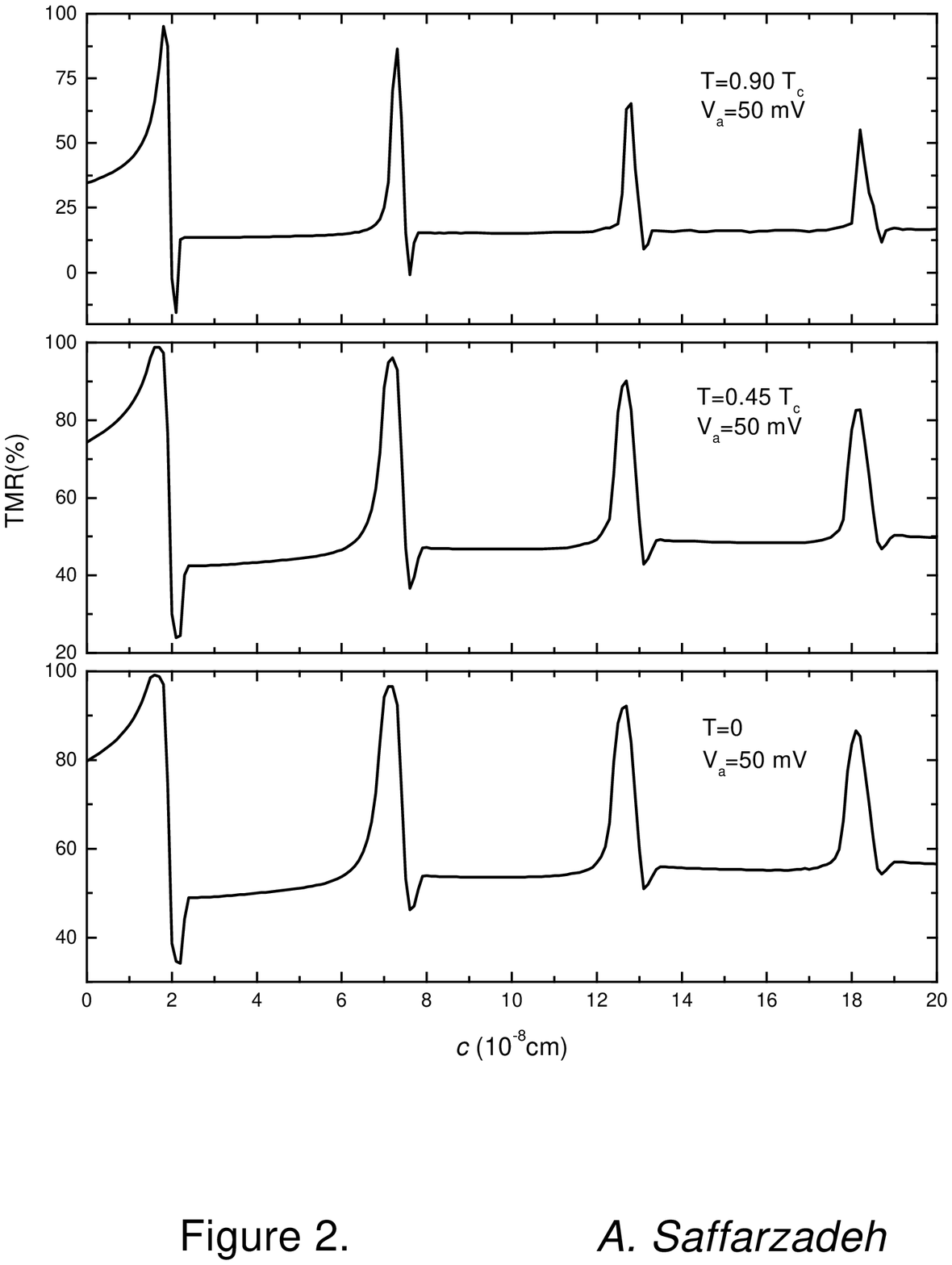}}
\end{center}
\caption{TMR as a function of the thickness $c$ in the EuS/NM double SF
junction at $T$=0, 0.45 and 0.9 $T_c$.}
\end{figure}
\newpage

\begin{figure}
\begin{center}
\leavevmode\hbox{\epsfxsize=1.1\textwidth\epsffile{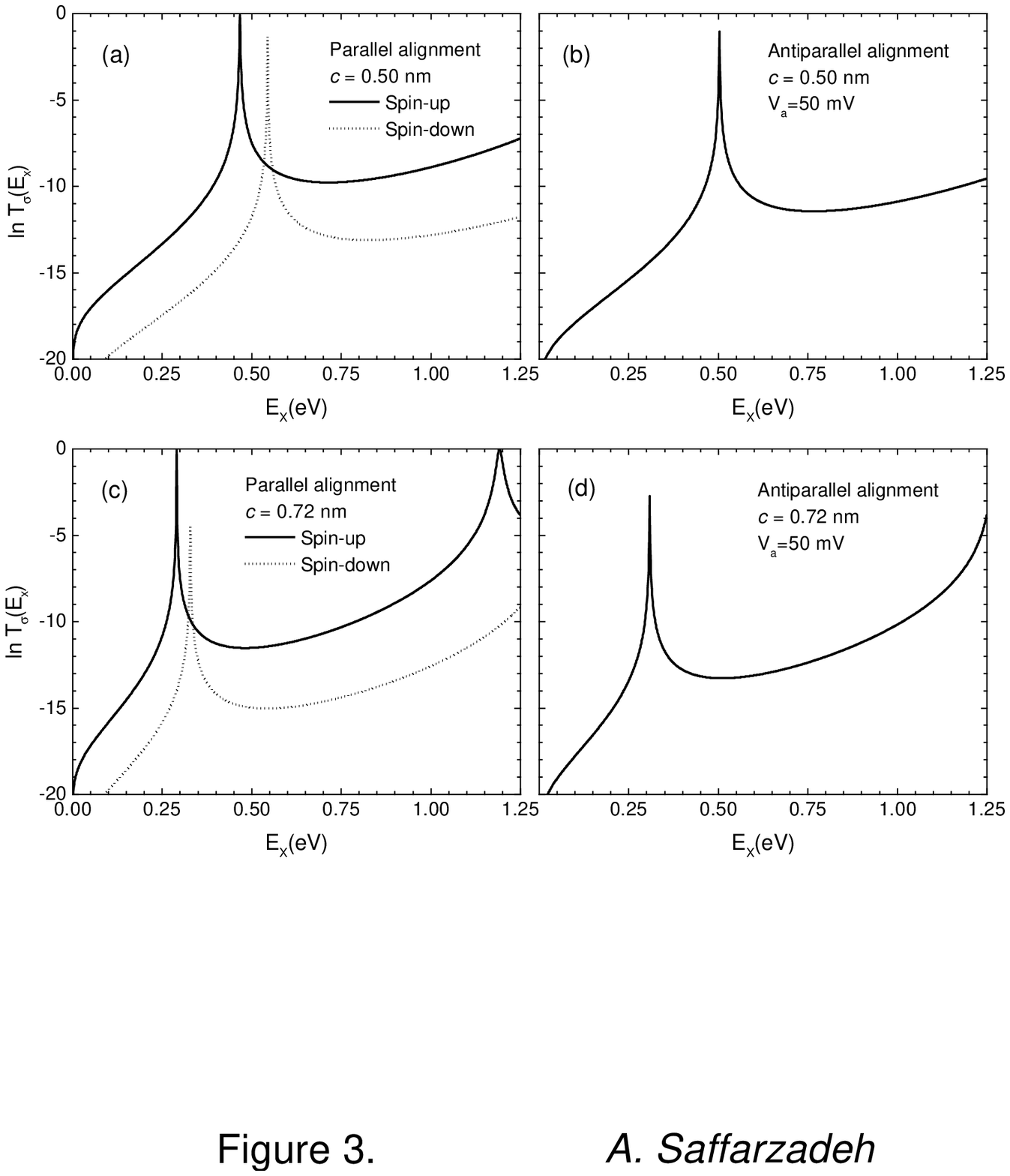}}
\end{center}
\caption{Spin-dependent transmission coefficients, $\ln T_{\s}(E_x)$, as a
function of energy $E_x$ in the EuS/NM double SF junction for $c$=0.50 and
0.72 nm at $T$=0 K.}
\end{figure}
\newpage

\begin{figure}
\begin{center}
\leavevmode\hbox{\epsfxsize=1.1\textwidth\epsffile{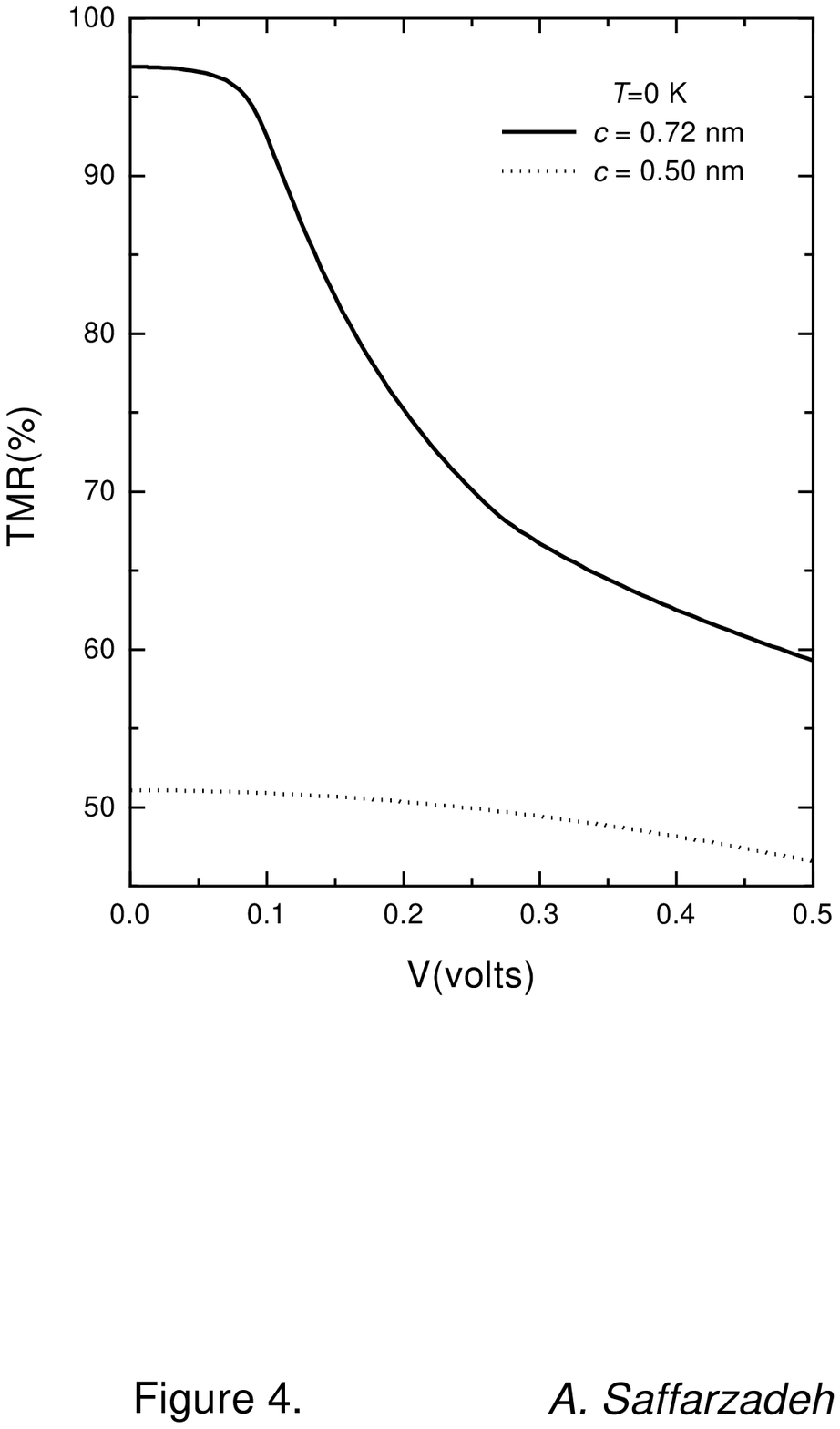}}
\end{center}
\caption{TMR as a function of applied bias $V_a$ in the EuS/NM double SF
junction for $c$=0.50 and 0.72 nm at $T$=0 K.}
\end{figure}
\newpage

\begin{figure}
\begin{center}
\leavevmode\hbox{\epsfxsize=1.1\textwidth\epsffile{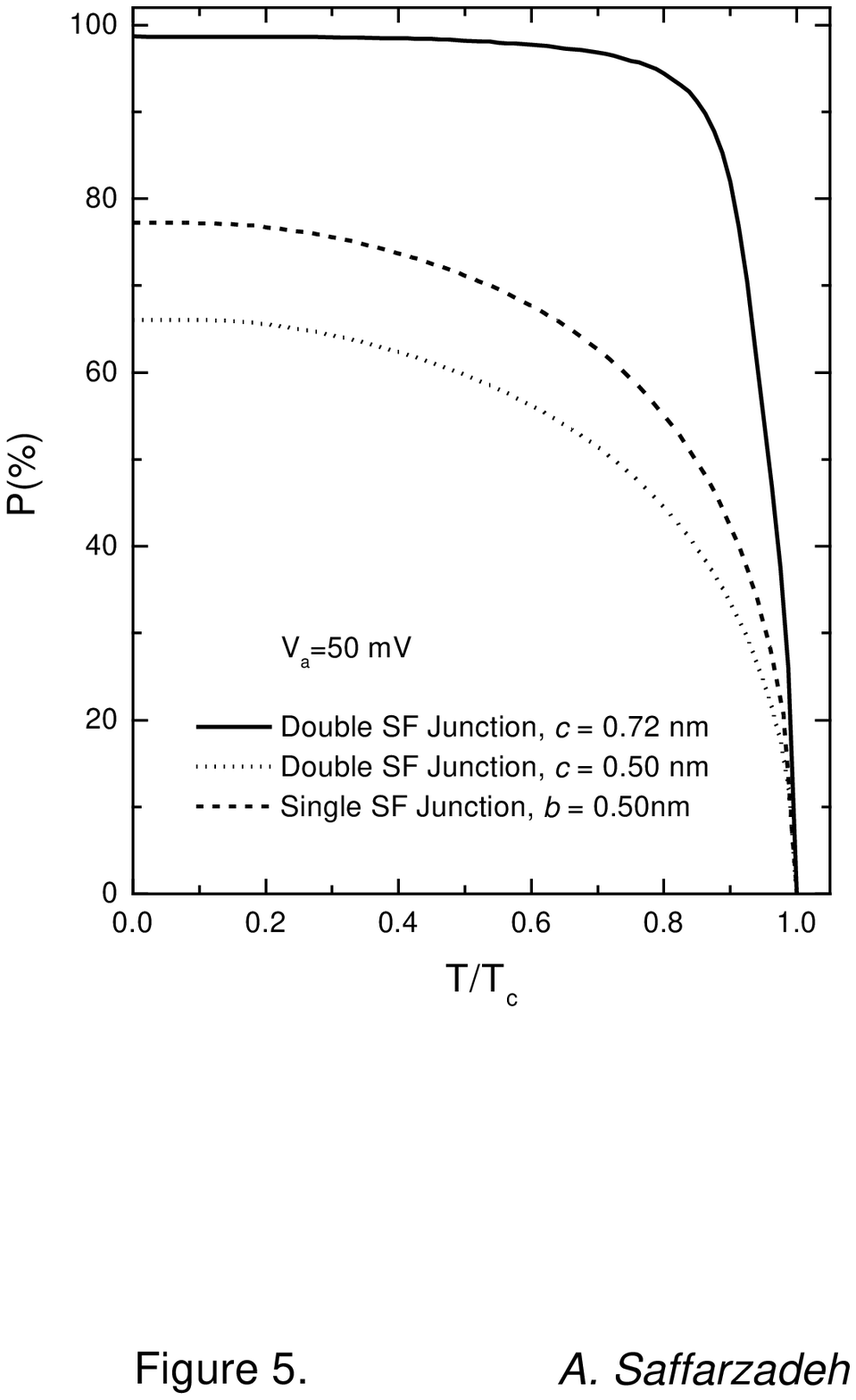}}
\end{center}
\caption{Electron-spin polarization $P$ as a function of normalized
temperature $T/T_C$ for the EuS/NM single and double SF junctions.}
\end{figure}

\end{document}